\documentclass[%
 reprint,
 amsmath,amssymb,
 aps,
 showkeys,
]{revtex4-2}

\usepackage{braket}
\usepackage{amsfonts}
\usepackage{color}
\usepackage[dvipsnames]{xcolor}
\usepackage{mathrsfs}
\usepackage{lmodern}
\usepackage{lineno}
\usepackage{hyperref}
\usepackage[normalem]{ulem}
\usepackage{verbatim} 
\expandafter\let\csname equation*\endcsname\relax
\expandafter\let\csname endequation*\endcsname\relax
\usepackage{amsmath}
\usepackage{graphicx}
\usepackage{dcolumn}
\usepackage{bm}

\newcommand{\beq}{\begin{eqnarray}}
\newcommand{\eeq}{\end{eqnarray}}
\newcommand{\nn}{\nonumber}
\DeclareMathOperator{\csch}{csch}
\newcommand{\tr}{\mathrm{tr}}

\begin{document}

\preprint{}

\title{Phase space quantization of anisotropic cosmologies: Taub and Kantowski-Sachs models.}

\author{Jasel Berra-Montiel}
\email{jasel.berra@uaslp.mx}
\author{Alberto Molgado}%
\email{alberto.molgado@uaslp.mx}
\author{Jorge Santacruz}
 \email{a331490@alumnos.uaslp.mx}
\affiliation{%
 Facultad de Ciencias, Universidad Aut\'onoma de San Luis 
Potos\'{\i} \\
Campus Pedregal, Av. Parque Chapultepec 1610, Col. Privadas del Pedregal, San
Luis Potos\'{\i}, SLP, 78217, Mexico \\
}%


\begin{abstract} 
We introduce an explicit construction of the non-diagonal and diagonal Wigner distributions for the homogeneous but anisotropic Taub and Kantowski-Sachs cosmological models within the framework of phase space deformation quantization. Conventional canonical quantization of these models via the Wheeler-DeWitt equation is inherently plagued by factor ordering ambiguities. To circumvent these issues, we employ the totally symmetric Weyl quantization map and the Moyal star product. By means of  a canonical separation of the Hamiltonian constraint, we are able to resolve the formal convergence problems typically associated with the star product. Furthermore, to establish a rigorous connection with conventional quantum cosmology, we calculate the standard wave functions directly from the diagonal Wigner distributions, recovering the exact physical states in terms of modified Bessel functions in both cases.
\end{abstract}

\keywords{Deformation quantization, Star product, Anisotropic Cosmology, Wigner distribution}

\maketitle

\section{Introduction}  
Deformation quantization also called phase space quantization introduces a well-defined framework for quantum mechanics within the classical phase space by replacing standard pointwise multiplication and Poisson brackets with a noncommutative, associative algebraic structure, commonly known as the Moyal star product~\cite{Bayen}, \cite{Curtright}, \cite{Teaching}, \cite{Cosmas}. It is formulated through a convolution procedure based on the Weyl quantization map, which provides a rigorous correspondence between classical phase space observables and totally symmetric operators defined on a Hilbert space. Besides using the star product it is possible to describe the temporal evolution of quantum systems by generalizing the classical Liouville equation, which governs the dynamics of the Wigner distribution, that is, the phase space quasi-probability analogue to the quantum density matrix. Historically, this robust mathematical framework has proven highly effective across various domains of physics, finding successful applications in quantum optics \cite{Schleich}, \cite{Hillery}, statistical mechanics \cite{Wigner}, \cite{Moyal}, and noncommutative field theories \cite{Douglas}, \cite{Seiberg}.
\\
\\
While standard cosmology heavily relies on the isotropic metric, understanding the dynamic of the primordial universe near the initial singularity requires examining homogeneous but anisotropic geometries \cite{Misner1969}. Among these, the Taub and Kantowski-Sachs universes stand out as fundamental minisuperspace models. The Taub universe is an exact vacuum solution with a closed $S^3$ topology, representing a highly symmetric, restricted case of the Bianchi IX classification \cite[p.~133]{RyanShepley1975}. Conversely, the Kantowski-Sachs model exhibits an $\mathbb{R} \times S^2$ topology and is formally isomorphic to the spacetime geometry inside the event horizon of a Schwarzschild black hole \cite[p.~168]{RyanShepley1975}, \cite{Kantowski1966}. From a quantum cosmological perspective, this isomorphism establishes a highly significant theoretical connection, implying that the interior of a black hole can be mathematically treated as a highly anisotropic, dynamical minisuperspace. Consequently, studying this model within the framework of deformation quantization allows for a simultaneous phase space characterization of both anisotropic cosmological scenarios and the quantum state of the black hole interior. Traditionally, the quantum behavior of these cosmological models has been explored using the canonical Wheeler-DeWitt equation \cite{DeWitt}. However, this framework often has problems with the inherent factor ordering ambiguities of noncommutative metric operators \cite{Halliwell}. Further, anisotropy in cosmology has been largely studied within the 
non-commutative phase space, as for example,~\cite{Vakili}, \cite{Barbosa}, \cite{Obregon} \cite{Nuno}, where they introduce noncommuting minisuperspace variables to study their effects on cosmological evolution and symmetries.  However, our focus is fundamentally different as we are able to  complete the quantization deformation program for the cosmological models of our interest. In consequence, our aim is to establish a precise method to calculate expectation values through a clear quasi-probabilistic perspective in phase space. Specifically, by calculating the Wigner distribution via deformation quantization for the Taub and the Kantowski-Sachs models, we can systematically control the mentioned operator ordering issues thanks to the totally symmetric nature inherent to the Weyl quantization map \cite[p.~79]{Folland}.
\\
\\
\indent The paper is organized as follows. In Section 2, we briefly review the concept of the Wigner distribution and the star product within the Deformation Quantization in phase space. In Section 3, we review both anisotropic cosmologies, the Taub and the Kantowski-Sachs models. In Section 4, we calculate the non-diagonal and diagonal Wigner distributions for both cosmological models. In Section 5, we establish the relation with conventional quantum mechanics calculating the wave functions for both models via its corresponding diagonal Wigner function. Finally, in Section 6, we present some concluding remarks and perspectives. Additionally, we present two brief appendices with the necessary calculations to support the results on Meijer G-functions for this work. 

\section{Deformation quantization}       
In this section, we outline the framework of deformation quantization, which provides the theoretical background required to formulate the quantum eigenvalue equations for the system in phase space. Here, we follow,  as close as possible, the notation in reference~\cite{star}. Broadly speaking, quantization stands for an arbitrary mathematical procedure that connect a classical mechanical system (typically governed by a Poisson geometry on a phase space $\Gamma$) with its quantum counterpart, represented by an algebra of self-adjoint operators acting on a Hilbert space $\mathcal{H}$. This transition is achieved through an invertible quantization map $\mathcal{Q}:f\mapsto\mathcal{Q}(f)$, which translates real-valued classical observables $f$ into these operators."
\\
\\
Our approach relies specifically on the Weyl quantization map. A significant advantage of this scheme is that it provides totally symmetric operators, therefore circumventing the well-known operator ordering ambiguities. For an observable $f(\mathbf{x},\mathbf{p})\in L^{2}(\mathbb{R}^{2n})$ defined on the classical phase space $\Gamma=\mathbb{R}^{2n}$, the corresponding Weyl operator acting on the Hilbert space $\mathcal{H}=L^{2}(\mathbb{R}^{n})$ is constructed via the Fourier transform $\widetilde{f}(\mathbf{a},\mathbf{b})$ as follows
\begin{equation}\label{Weyl}
	\mathcal{Q}_W(f)=
	\frac{1}{(2\pi)^{n}}
	\int_{\mathbb{R}^{2n}}
	\widetilde{f}(\mathbf{a},\mathbf{b})
	e^{i(\mathbf{a}\cdot\mathbf{\widehat{X}}+\mathbf{b}\cdot\mathbf{\widehat{P}})}
	d\mathbf{a}\,d\mathbf{b}\,.
\end{equation}
Here, the position and momentum operator components satisfy the canonical commutation algebra $[\widehat{X}_{i},\widehat{P}_{j}]=i\hbar\delta_{ij}$. As discussed in \cite{Takhtajan}, the integral in (\ref{Weyl}) does not necessarily converge absolutely and is generally understood within the weak operator topology. The action of this operator on a generic quantum state $\psi\in L^{2}(\mathbb{R}^{n})$ can be expressed through an integral kernel $\kappa_{f}(\mathbf{x},\mathbf{y})\in L^2(\mathbb{R}^{2n})$, ensuring $\mathcal{Q}_{W}(f)\equiv\widehat{F}$ is a Hilbert-Schmidt operator \cite{Reed}
\begin{widetext}
\begin{equation}
		\mathcal{Q}_{W}(f)\psi(\mathbf{x})=
		\, \frac{1}{(2\pi\hbar)^{2n}}
		\int_{\mathbb{R}^{2n}}
		f\left(\frac{\mathbf{x}+\mathbf{y}}{2},\mathbf{p}\right)
		e^{-\frac{i}{\hbar}(\mathbf{y}-\mathbf{x})\cdot\mathbf{p}}
		\psi(\mathbf{y})
		d\mathbf{p}\,d\mathbf{y}
		=\int_{\mathbb{R}^{n}}
		\kappa_{f}(\mathbf{x},\mathbf{y})
		\psi(\mathbf{y})
		d\mathbf{y}\,.
\end{equation}
\end{widetext}
Conversely, one can map a quantum operator $\widehat{F}\in HS(L^{2}(\mathbb{R}^{n}))$ back to its classical phase space representation. This inverse Weyl map yields what is known in harmonic analysis \cite{Reed}, \cite{Folland} as the \textit{symbol} $f$ corresponding to the operator $\widehat{F}$, defined by $f(\mathbf{x},\mathbf{p})=\tr(\widehat{\Omega}(\mathbf{x},\mathbf{p})\widehat{F})$, which explicitly reads
\begin{equation}\label{InverseWeyl}
	\mathcal{Q}_{W}^{-1}(\widehat{F})(\mathbf{x},\mathbf{p})=
	\hbar^{n}
	\int_{\mathbb{R}^{n}}
	\kappa_{f}\left(\mathbf{x}-\frac{\hbar\mathbf{y}}{2},\mathbf{x}+\frac{\hbar\mathbf{y}}{2}\right)
	e^{i\mathbf{y}\cdot\mathbf{p}}
	d\mathbf{y}\,,
\end{equation}
where the operator $\widehat{\Omega}(\mathbf{x},\mathbf{p})$ corresponds to the Weyl-Stratonovich operator. While this framework naturally extends to the space of tempered distributons $\mathcal{S}'(\mathbb{R}^{2n})$, we restrict our current scope to square-integrable functions (see \cite{Folland} for an exhaustive treatment from the point of view of analysis). Applying this inverse mapping to the density matrix operator allows us to construct the Wigner distribution. If we consider a pure quantum state $\psi\in L^{2}(\mathbb{R}^{n})$ with its associated density operator $\widehat{\rho}=\ket{\psi}\bra{\psi}$, the application of the inversion formula (\ref{InverseWeyl}) provides its phase space symbol
\beq
W(\mathbf{x},\mathbf{p})
	& = &
    Q_{W}^{-1}(\widehat{\rho})(\mathbf{x},\mathbf{p}) \nonumber\\
    & = &
	\frac{1}{(2\pi\hbar)^{n}} 
	\int_{\mathbb{R}^{n}}
	\psi\left( \mathbf{x}+\frac{\mathbf{y}}{2}\right)
	\psi^{*}\left( \mathbf{x}-\frac{\mathbf{y}}{2}\right) e^{-\frac{i}{\hbar}\mathbf{y}\cdot\mathbf{p}}
	d\mathbf{y}\,.  \nonumber\\
	&  & 
\label{WD-WF}
\eeq
This representation of the density operator is defined as the Wigner distribution. Wigner distribution enables the study of quantum systems using probability distributions over phase space, closely mirroring classical statistical mechanics \cite{Wigner}. Wigner distribution is called a quasi-probability distribution because, unlike classical probability densities, it can take negative values. Nonetheless, it perfectly facilitates the calculation of quantum expectation values by weighting the symbol $A = \mathcal{Q}^{-1}_{W}(\widehat{A})$ over the entire phase space
\begin{equation}
	\braket{\psi|\widehat{A}\psi}
	=\int_{\mathbb{R}^{2n}}
	W(\mathbf{x},\mathbf{p})
	A(\mathbf{x},\mathbf{p})d\mathbf{x}
	d\mathbf{p}\,.
\end{equation}  
\\
We turn now to the definition of the star product. Given that the Weyl map is a bijection onto the space of Hilbert-Schmidt operators and since the product of such operators is closed \cite{Folland}, the standard operator composition $\mathcal{Q}_{W}(f)\mathcal{Q}_{W}(g)$ yields another valid operator in the Hilbert space $\mathcal{H}$. Consequently, the bijectivity implies there exists a unique function on $L^{2}(\Gamma)$, where $\Gamma=\mathbb{R}^{2n}$ is the classical phase space, whose Weyl map corresponds exactly to this operator product. We denote this unique function by $f\star g$, which implicitly defines the star product through the relation $\mathcal{Q}_{W}(f\star g)=\mathcal{Q}_{W}(f)\mathcal{Q}_{W}(g)$. By utilizing (\ref{Weyl}) and (\ref{InverseWeyl}) and evaluating the exponents, one arrives at its integral representation \cite{Cosmas}
\beq
	(f\star g)(\mathbf{x},\mathbf{p})
	& = &
	\frac{1}{(\pi\hbar)^{2}}
	\int_{\mathbb{R}^{4}} d\mathbf{a}\,d\mathbf{b}\,d\mathbf{c}\,d\mathbf{d}  \,
	f(\mathbf{a},\mathbf{b})
	g(\mathbf{c},\mathbf{d}) \nonumber  \\
	& & \times
	\left[e^{-\frac{2i}{\hbar}(\mathbf{p}\cdot(\mathbf{a}-\mathbf{c})+\mathbf{x}\cdot(\mathbf{d}-\mathbf{b})+(\mathbf{c}\cdot\mathbf{b}-\mathbf{a}\cdot\mathbf{d}))}\right] \!.
\label{SP-Int}
\eeq  
For functions $f,g\in C^{\infty}(\mathbb{R}^{2n})$, a Taylor expansion of the integral kernel in (\ref{SP-Int}) yields the widely used pseudo-differential form of the star product \cite{Teaching}
\begin{widetext}
\begin{equation}
		(f\star g)(\mathbf{x},\mathbf{p}) := f(\mathbf{x},\mathbf{p})
		\exp\left[\frac{i\hbar}{2}\left(\frac{\overleftarrow{\partial}}{\partial \mathbf{x}}\cdot\frac{\overrightarrow{\partial}}{\partial\mathbf{p}}- \frac{\overleftarrow{\partial}}{\partial \mathbf{p}}\cdot\frac{\overrightarrow{\partial}}{\partial\mathbf{x}}\right)\right]
		g(\mathbf{x},\mathbf{p})  
		= \sum_{m,n=0}^{\infty}
		\left( \frac{i\hbar}{2}\right)^{m+n}
		\frac{(-1)^{m}}{m!n!}
		\left( \partial^{m}_{\mathbf{p}}\cdot\partial^{n}_{\mathbf{x}}f\right)
		\left( \partial^{n}_{\mathbf{p}}\cdot\partial^{m}_{\mathbf{x}}g\right)\,,
\end{equation} 
\end{widetext}  
where the arrows in the first identity indicate the direction in which the differential operators act. Notice that the pointwise product is recovered in the classical limit $\hbar\to 0$. This star product naturally leads to the Moyal bracket
\begin{equation}
    \left\lbrace f,g\right\rbrace_{M}=(i\hbar)^{-1}\left(f\star g-g\star f\right)\,,
\end{equation}
which stands for a quantum deformation of the classical Poisson bracket. Consequently, Weyl's correspondence transforms the quantum von Neumann equation into the Moyal's evolution equation \cite{Curtright}
\begin{equation}
	\frac{\partial W(\mathbf{x},\mathbf{p})}{\partial t}=
	\left\lbrace W,H\right\rbrace_{M}\,,  
\end{equation}
dictating the dynamics of the Wigner distribution, with $H$ being the Hamiltonian symbol. Like in he standard Schrödinger picture, spectral properties within deformation quantization arise from star-eigenvalue problems. For a stationary quantum state, the energy spectrum is extracted via $H(\mathbf{x},\mathbf{p})\star W(\mathbf{x},\mathbf{p})=E W(\mathbf{x},\mathbf{p})$. Generalizing to consider non-stationary transition Wigner distributions (non-diagonal Wigner distribution), perhaps it will be $W_{E_{L}E_{R}}(\mathbf{x},\mathbf{p})$, the formulation requires a coupled system of equations corresponding to the left and the right star-multiplications
\begin{equation}\label{Non-diagonal Wigner}
	\begin{split}
		H(\mathbf{x},\mathbf{p})\star W_{E_{L}E_{R}}(\mathbf{x},\mathbf{p})=&
		E_{L}\,
		W_{E_{L}E_{R}}(\mathbf{x},\mathbf{p})\,,\\
		W_{E_{L}E_{R}}(\mathbf{x},\mathbf{p})\star H(\mathbf{x},\mathbf{p})=&
		E_{R}\,
		W_{E_{L}E_{R}}(\mathbf{x},\mathbf{p})\,,
	\end{split}
\end{equation}
where $E_{L}$ and $E_{R}$ represent the energy eigenvalue of the transition, respectively,~\cite{Curtright},~\cite{Morse}. Non-diagonal Wigner distribution is the symbol of the matrix density $\widehat{\rho}_{E_{L}E_{R}}=|E_{L}\rangle \langle E_{R}|$, and the general matrix density $\widehat{\rho}(t)$ can be constructed via a linear combination of it.

\section{The Taub and Kantowski-Sachs cosmological models}
The description of gravitational dynamics in the primordial stages of the universe may require relaxing the strict isotropy principle of the Friedmann-Lema\^itre-Robertson-Walker (FLRW) metric to introduce homogeneous but anisotropic models \cite{Misner1969}. This theoretical framework is rigorously established using the Arnowitt-Deser-Misner (ADM) formalism, which foliates the four-dimensional curved spacetime into a family of three-dimensional spatial hypersurfaces $\Sigma_{t}$ parameterized by a temporal coordinate \cite{Kiefer}. Following a Legendre transformation, the gravitational vacuum dynamics adopts the structure of a parameterized Hamiltonian system, governed primarily by the constraint $H\approx 0$ \cite{DeWitt}. The ADM formalism demonstrates that General Relativity, when is formulated in a $3+1$ spacetime decomposition, possesses two physical degrees of freedom per spatial point, thus constituting a field theory with an infinite number of degrees of freedom. In order to provide a quantization process mathematically tractable, this full configuration space traditionally known as superspace is truncated to a finite-dimensional subspace called minisuperspace by imposing strict spatial symmetries. Specifically, requiring spatial homogeneity confines the infinite gravitational degrees of freedom to a finite set of time-dependent scale factor and anisotropy parameters. In particular, in the cosmological setup, a systematic framework for constructing such spatially homogeneous spacetimes is provided by the Bianchi classification, which categorizes all three-dimensional real Lie algebras according to their structure constants~\cite{RyanShepley1975}. These algebras dictate the commutation relations of the Killing vector fields that generate a isometry groups acting simply transitively on the spatial hypersurfaces, thus offering a rigorous mathematical foundation to study homogeneous but anisotropic cosmological models \cite{RyanShepley1975}. Some other interesting minisuperspaces are given by the Taub and the Kantowski-Sachs models that we are considering in this work, where the Taub model represents a special case on Bianchi classification. Applying näive Dirac's canonical quantization promotes the metric variables and their conjugate momenta to  operators and consequently, the classical constraint is transformed into the fundamental Wheeler-DeWitt equation \cite{DeWitt, Halliwell}
\begin{equation}\label{WD}
	\hat{H}\psi=0,
\end{equation}
where $\psi$ is the wave function of the system.

\subsection{Classical and Quantum Taub Universe}
The Taub universe represents an exact cosmological vacuum solution with a closed $S^{3}$ topology, mathematically categorized as a special case of the Bianchi Type IX geometry \cite[p.~168]{RyanShepley1975}. This specific configuration is recovered by imposing a Locally Rotationally Symmetric (LRS) condition, which permanently constrains one of the anisotropic degrees of freedom ($\beta_{-}=0$) throughout its evolution \cite{Misner1969}, \cite{Ellis}. Adopting Misner's parameterization, the kinematics of the space are described by an isotropic volume factor $\Omega$ and the residual shear parameter $\beta_{+}$. The reduced Hamiltonian constraint for the classical Taub model is given by
\begin{equation}\label{TaubH}
	H_{T}=
	\pi_{+}^2-
	\pi_{\Omega}^{2}+
	\frac{1}{3}e^{4\Omega}(e^{-8\beta_{+}}-
	4e^{-2\beta_{+}}) \approx 0\,,
\end{equation}
where $\pi_{\Omega}$ and $\pi_{+}$ are the conjugate momenta associated with the volume and the anisotropy respectively \cite[p.~68]{Sime}. The intrinsic spatial curvature acts as an effective, exponentially repulsive potential barrier \cite{Misner2}. Classically, the system behaves analogously to a particle in the configuration space that undergoes a finite number of elastic bounces against this barrier before inexorably collapsing into an initial cosmological singularity with divergent curvature \cite{Misner1969}, \cite{RyanShepley1975}. Within quantum cosmology, canonical quantization via the Wheeler-DeWitt equation promotes this classical system to a quantum wave dynamic, which inherently retains the curvature-induced potential walls of the classical minisuperspace \cite{Misner2}. Specifically, at the quantum level exact analytical solutions to this system manifest as a product of two modified Bessel functions, mathematically confirming the absolute confinement of the wave packet against the exponential anisotropy barriers in the reduced configuration space \cite{Mon}, \cite{Sime1}. However, rigorous analyses using asymptotic semiclassical regimes (WKB approximation) have demonstrated that standard canonical quantization is insufficient to fully regularize the model \cite{Cascioli}. The wave packet dispersion caused by interference against the anisotropy barrier fails to halt the probabilistic trajectory towards the zero-volume convergence, meaning the standard Wheeler-DeWitt formulation alone cannot completely eradicate the initial singularity \cite{Misner2}, \cite{Battisti}.

\subsection{Classical and Quantum Kantowski-Sachs Model}
In contrast to the closed spherical topology of the Taub model, the Kantowski-Sachs universe exhibits a product structure $\mathbb{R} \times S^{2}$ \cite{Kantowski1966}. As it admits a continuous four-parameter isometry group acting with multiple degrees of transitivity on the spatial hypersurfaces, this space belongs exclusively to the LRS category and falls outside the standard Bianchi classification \cite[p.~168]{RyanShepley1975}, \cite{Ellis}. Its primary phenomenological significance lies in the exact analytical isomorphism, that may be established between this model and the confined interior geometry of a Schwarzschild black hole horizon ($r<2GM$). Inside this dynamically homogeneous region, the radial gradient vector transforms into a temporal one, forcing the orthogonal hypersurfaces to undergo a directionally asymmetric collapse \cite{Kantowski1966}, \cite[p.~819]{Thorne}. Analogously adopting Misner's parameterization, the kinematics of this space are governed by the longitudinal and transversal scale factors, represented by an isotropic volume factor $\Omega$ and a single anisotropic shear parameter $\beta$. The Hamiltonian constraint for the Kantowski-Sachs model takes the form
\begin{equation}\label{HSH}
	H_{KS}=-
	\pi_{\Omega}^{2}+
	\pi_{\beta}^{2}-
	e^{4\Omega+2\beta} \approx 0\,,
\end{equation}
with $\pi_{\Omega}$ and $\pi_{\beta}$ being the conjugate momenta associated with the volume and the anisotropy respectively \cite[p.~63]{Sime}. Classically, this metric evolution leads to a directional singularity, acting as a terminal temporal spacelike singularity \cite{Kantowski1966}. Upon quantizing the Kantowski-Sachs geometry via the Wheeler-DeWitt equation, exact wave function solutions can be constructed using expansions of modified Bessel functions, depending on the chosen algebraic prescription \cite{Sime2}, \cite{Cordero}. While these solutions help stabilize the probability distribution against invariants that diverge classically, canonical quantum cosmology still struggles to definitively resolve the singularity without invoking discrete spacetime structures \cite{Misner2}, \cite{Battisti}.
\\
\\
\indent As before mentioned, the standard Wheeler-DeWitt formulation, suffers from factor ordering ambiguities due to the nature of noncommutative metric operators. 
However, the phase space deformation quantization framework provides a systematic approach to address this issue by employing the Moyal star product, that is, the formalism canonically selects the totally symmetric ordering prescription via the Weyl correspondence which maps the classical cosmological Hamiltonian to a fully symmetrized quantum operator. Consequently, this approach yields a well-defined Moyal-Wheeler-DeWitt equation within the Weyl quantization framework \cite{Rashki}
\begin{equation}\label{MWDW}
	H(\mathbf{x},\mathbf{p})\star W(\mathbf{x},\mathbf{p})=0\,.
\end{equation}

\section{Phase space quantization: Taub  and Kantowski-Sachs models}
A fundamental issue in formal deformation quantization is the convergence problem. Typically, the formal star product between general smooth functions leads to a power series with a strict zero radius of convergence. This divergent behavior is particularly severe for systems governed by exponential potentials \cite{Waldmann}. However, in the minisuperspaces studied here, the intrinsic symmetries of the geometry prevent such divergences. The absence of cross-terms in the exponential potentials allows for the complete canonical separation of the Hamiltonian constraints. This separability effectively reduces the multidimensional star product eigenvalue problem into independent, exactly solvable one-dimensional equations, ensuring the mathematical convergence of the deformed algebra. Consequently, the method we employ to find the corresponding Wigner distribution for each cosmological model is based on Belchev and Walton's framework \cite{Morse}. Where we apply the equations (\ref{Non-diagonal Wigner}) to obtain the non-diagonal Wigner distribution and furthermore, show the diagonal case. Before proceeding with the explicit calculations, it is important to contextualize the use of Deformation Quantization within the framework of noncommutative quantum cosmology. In this approach, phase space noncommutativity is typically introduced by imposing a deformed algebraic structure, often via a Moyal type product directly onto the minisuperspace variables. For example, in~\cite{Vakili} the authors explored the effects of phase space noncommutativity on the classical and quantum dynamics of Bianchi class A spacetimes, revealing how the deformed algebra modifies the universe's evolution and its underlying Noether symmetries. Similarly, in \cite{Obregon} it is  proposed a noncommutative deformation for the Kantowski-Sachs minisuperspace, demonstrating that this non-commutativity generates new probable quantum states for the early universe, which may be connected through tunneling processes. Furthermore, while previous works have calculated Wigner distributions for these cosmological scenarios indirectly from known Wheeler-DeWitt solutions \cite{Cordero}, our approach analytically solves the star eigenvalue equations directly in phase space without previous dependence on the standard wave functions.

\subsection{Wigner distribution for Taub Model}
In order to continue we use the Hamiltonian constraint for the Taub model from equation (\ref{TaubH}), then, we apply a canonical transformation, which allow us to separate the Hamiltonian constraint. We would like to introduce a linear transformation, since the star product remains invariant under linear canonical transformations \cite{Curtright}. Introducing the new coordinates $x=\Omega-2\beta_{+}$ and $y=2\Omega-\beta_{+}$, and employing the generating function $F_{2}$ \cite{Gold}. In this specific case, the generating function is $F_{2}(\Omega,\beta_{+},p_{x},p_{y})=(\Omega-2\beta_{+})p_{x}+(2\Omega-\beta_{+})p_{y}$, the original canonical momenta are expressed as $\pi_{\Omega}=p_{x}+2p_{y}$ and $\pi_{+}=-2p_{x}-p_{y}$. Conversely, the new momenta are given by $p_{x}=-\frac{1}{3}\left(\pi_{\Omega}+2\pi_{+}\right)$, and $p_{y}=\frac{1}{3}\left(2\pi_{\Omega}+\pi_{+}\right)$. This canonical transformation lead us to the Hamiltonian constraint in terms of those variables
\begin{equation}\label{Hamiltonian}
	H=
	p_{x}^{2}-
	p_{y}^{2}+
	\frac{1}{9}
	\left(e^{4x}-4e^{2y}\right)
	=H_{x}
	-H_{y}
	\approx 0\,,
\end{equation}
where $H_{x}=p_{x}^{2}+\frac{1}{9}e^{4x}$ and $H_{y}=p_{y}^{2}+\frac{4}{9}e^{2y}$. Since this new Hamiltonian is separable, its associate non-diagonal Wigner distribution is $W_{E_{Lxy}E_{Rxy}}(x,y,p_{x},p_{y})=W_{E_{Lx}E_{Rx}}(x,p_{x})W_{E_{Ly}E_{Ry}}(y,p_{y})$. This allow us to calculate each distribution separately and, as a consequence, we may correspondingly consider the eigenvalues $E_{Lx},\,\,E_{Ly}$ and $E_{Rx},\,\,E_{Ry}$ for each case, as both $W_{E_{Lx}E_{Rx}}(x,p_{x})$ and $W_{E_{Ly}E_{Ry}}(y,p_{y})$ posses its own energy basis. First of all, taking the Hamiltonian $H_{x}$ to calculate the star-eigenvalue equations system from equation (\ref{Non-diagonal Wigner}) for $W_{E_{Lx}E_{Rx}}(x,p_{x})$ which, by simplicity in the notation, we will write as $W$ in what follows. In particular we may consider the next equation
\begin{widetext}
\begin{equation}
		H_{x}\star W=
		\left(p_{x}^{2}+\frac{1}{9}e^{4x}\right)
		\exp\left[\frac{i\hbar}{2}\left(\frac{\overleftarrow{\partial}}{\partial x}\cdot\frac{\overrightarrow{\partial}}{\partial p_{x}}- \frac{\overleftarrow{\partial}}{\partial p_{x}}\cdot\frac{\overrightarrow{\partial}}{\partial x}\right)\right]
		W
		= \ p_{x}^{2}\,W-
		i\hbar p_{x}\frac{\partial W}{\partial x}-
		\frac{\hbar^{2}}{4}\frac{\partial^{2} W}{\partial x^{2}}+
		\frac{e^{4x}}{9}
		\sum_{n=0}^{\infty}
		\frac{(2i\hbar)^{n}}{n!}
		\frac{\partial^{n} W}{\partial p_{x}^{n}}\,.
\end{equation}
\end{widetext}
Using the Taylor expansion definition, we can rewrite the infinity sum as a displacement over the momenta of $W$, thus
\beq
	H_{x}\star W 
	& = &
	p_{x}^{2}\,W-
	i\hbar p_{x}\frac{\partial W}{\partial x}-
	\frac{\hbar^{2}}{4}\frac{\partial^{2} W}{\partial x^{2}}
	\nonumber\\
	& & +
	\frac{e^{4x}}{9}
	W(x,p_{x}+2i\hbar) = 
	E_{Lx}\,W\,.
	\label{ELx-EQ}
\eeq
Now we define $W_{E_{Lx}E_{Rx}}(x,p_{x})=f(u,p_{x})$, where $u=e^{-8x}$ and we used $f$ for simplicity. Rewriting equation (\ref{ELx-EQ}) yields
\beq
	p_{x}^{2}\, f+
	8i\hbar p_{x}u\frac{\partial f}{\partial u}-
	16\hbar^{2}\left(u\frac{\partial f}{\partial u}+u^{2}\frac{\partial^{2} f}{\partial u^{2}}\right)  \nonumber\\
	+
	\frac{u^{-1/2}}{9}
	f(u,p_{x}+2i\hbar)
	=E_{Lx}\, f\,.
	\label{f-EQ}
\eeq
Next, we use the Mellin transform \cite[p.~367]{Debnath}, defined as
\begin{equation}
    \mathcal{M}\{g\}(s):=\int_{0}^{\infty}z^{s-1}g(z)\,dz\,.
\end{equation}
Applying this transform to both sides of the equation (\ref{f-EQ}) yields to transform the differential equation into an algebraic equation, consequently simplifying its analysis
\beq
		(p_{x}-4i\hbar s)^{2}F(s,p_{x})+
		\frac{1}{9}F(s-1/2,p_{x}+2i\hbar)
		& = &  \nn\\
		E_{Lx}\,F(s,p_{x})\,,\nn\\
		(p_{x}+4i\hbar s)^{2}F(s,p_{x})+
		\frac{1}{9}F(s-1/2,p_{x}-2i\hbar)
		& = &  \nn\\
		E_{Rx}\,F(s,p_{x})\,,
\label{Mellin-EQ}
\eeq
where after apply the Mellin transform and grouping terms we obtain the first equation from (\ref{Mellin-EQ}). Second equation is obtained analogously via the equation $W\star H_{x}=E_{Rx}\,W$ and $F(s,p_{x})=\mathcal{M}\{f\}(s,p_{x})$ is the Mellin transform in the first entry of $f(u,p_{x})$. Furthermore, we assume that $F$ is separable in terms of $E_{Lx}$ and $E_{Rx}$
\begin{equation}
	F(s,p_{x})=
	\mathcal{N}
	F_{Lx}\left(s+\frac{ip_{x}}{4\hbar},E_{Lx}\right)
	F_{Rx}\left(s-\frac{ip_{x}}{4\hbar},E_{Rx}\right)\,,
\end{equation}
where $\mathcal{N}$ is a normalization constant. Now, we can calculate $F_{Lx}$ and $F_{Rx}$ separately using its corresponding recurrence relation from equation (\ref{Mellin-EQ}). Taking $F_{Lx}$ and defining $t=s+ip_{x}/4\hbar$, and equivalently, taking $F_{Rx}$ and defining $\bar{t}=s-ip_{x}/4\hbar$, we obtain two algebraically identical equations
\beq
		F_{Lx}(t-1,E_{Lx})
		& = &  
		144\hbar^{2}
		\left(\frac{i\sqrt{E_{Lx}}}{4\hbar}-t\right)
		\left(-\frac{i\sqrt{E_{Lx}}}{4\hbar}-t\right)  \nn\\
		& & \times
		F_{Lx}(t,E_{Lx})\,,\nn\\
		F_{Rx}(\bar{t}-1,E_{Rx})
		& = &
		144\hbar^{2}
		\left(\frac{i\sqrt{E_{Rx}}}{4\hbar}-\bar{t}\right)
		\left(-\frac{i\sqrt{E_{Rx}}}{4\hbar}-\bar{t}\right) \nn\\
		& & \times
		F_{Rx}(\bar{t},E_{Rx})\,.
\eeq
Thus, developing both recurrence relations yields the followings solutions in terms of the Gamma functions $\Gamma(z)$~\cite{G}
\begin{equation}
	\begin{split}
		F_{Lx}\left(t,E_{Lx}\right)=&
		\frac{1}{(144\hbar^{2})^{t}}
		\Gamma\left[-t+\frac{i\sqrt{E_{Lx}}}{4\hbar}\right]
		\Gamma\left[-t-\frac{i\sqrt{E_{Lx}}}{4\hbar}\right],\\
		F_{Rx}\left(\bar{t},E_{Lx}\right)=&
		\frac{1}{(144\hbar^{2})^{\bar{t}}}
		\Gamma\left[-\bar{t}+\frac{i\sqrt{E_{Rx}}}{4\hbar}\right]
		\Gamma\left[-\bar{t}-\frac{i\sqrt{E_{Rx}}}{4\hbar}\right].
	\end{split}
\end{equation}
Now, we can write the complete solution for $F(s,p_{x})$ replacing $t$ and $\bar{t}$ as follows
\begin{align}
	F(s,p_{x})=&
	\mathcal{N}(144\hbar^{2})^{-2s}\,
	\Gamma\left[-s+\frac{i}{4\hbar}\left(-p_{x}+\sqrt{E_{Lx}}\right)\right]\nonumber\\
	&\times\Gamma\left[-s+\frac{i}{4\hbar}\left(-p_{x}-\sqrt{E_{Lx}}\right)\right]  \nn\\
	&\times
	\Gamma\left[-s+\frac{i}{4\hbar}\left(p_{x}+\sqrt{E_{Rx}}\right)\right]\nonumber\\
	&\times\Gamma\left[-s+\frac{i}{4\hbar}\left(p_{x}-\sqrt{E_{Rx}}\right)\right]\,.
\end{align}
In order to continue, we must take this solution and apply the inverse Mellin transform \cite{Debnath} to return to our original space, such inverse transform is defined as
\begin{equation}
    \mathcal{M}^{-1}\{G\}(z):=\frac{1}{2\pi i}\int_{c-i\infty}^{c+i\infty}z^{-s}G(s)\,ds,
\end{equation}
where $c$ is a complex number whose real part need satisfy a mild lower bound. This inverse Mellin transform lead us to the integral
\begin{align}
	f(u,p_{x})=&
	\frac{\mathcal{N}}{2\pi i}
	\int_{-i\infty}^{i\infty} ds\,
	\Gamma\left[-s+\frac{i}{4\hbar}\left(-p_{x}+\sqrt{E_{Lx}}\right)\right]\nonumber\\
	&\times\Gamma\left[-s+\frac{i}{4\hbar}\left(-p_{x}-\sqrt{E_{Lx}}\right)\right]  \nn\\
	&\times
	\Gamma\left[-s+\frac{i}{4\hbar}\left(p_{x}+\sqrt{E_{Rx}}\right)\right]\nonumber\\
	&\times\Gamma\left[-s+\frac{i}{4\hbar}\left(p_{x}-\sqrt{E_{Rx}}\right)\right]
	((144\hbar^{2})^{2}u)^{-s}\,.
\end{align}
This integral may be recognized as the Meijer G-function $G_{0,4}^{4,0}$ definition via the Mellin–Barnes integral \cite[P.~1041]{G}. Replacing $u$, we obtain the solution
\begin{widetext}
\begin{equation}\label{Meijerx-Taub}
	W_{E_{Lx}E_{Rx}}(x,p_{x})=
	\mathcal{N}\,
	G_{0,4}^{4,0}
	\left(\frac{e^{8x}}{(12\hbar)^{4}}\ 
	\middle|
	\frac{i}{4\hbar}\left(-p_{x}\pm\sqrt{E_{Lx}}\right), \frac{i}{4\hbar}\left(p_{x}\pm\sqrt{E_{Rx}}\right)\right).
\end{equation}
\end{widetext}
Proceeding in a complete analogous manner, in order to calculate $W_{E_{Ly}E_{Ry}}(y,p_{y})$, we must apply identical arguments as the previous case $W_{E_{Lx}E_{Rx}}(x,p_{x})$, except that now we use the Hamiltonian $\mathcal{H}_{y}$, which lead us to
\begin{widetext}
\begin{equation}\label{Meijery-Taub}
	W_{E_{Ly}E_{Ry}}(y,p_{y})=
	\mathcal{K}\,
	G_{0,4}^{4,0}
	\left(\frac{e^{4y}}{(3\hbar)^{4}}\
	\middle|
	\frac{i}{2\hbar}\left(-p_{y}\pm\sqrt{E_{Ly}}\right), \frac{i}{2\hbar}\left(p_{y}\pm\sqrt{E_{Ry}}\right)\right)\,,
\end{equation}
\end{widetext}
where $\mathcal{K}$ is its own normalization constant. Since we have the Hamiltonian constraint $H=H_{x}-H_{y}\approx 0$, it must follow the Moyal-Wheeler-DeWitt equation (\ref{MWDW}). In the particular case of our interest, it then must hold
\begin{equation}\label{M-W-D}
	\begin{split}
		H\star W=&
		H_{x}\star W-
		H_{y}\star W=0\,,\\
		W\star H=&
		W\star H_{x}-W\star H_{y}=0\,.
	\end{split}
\end{equation}
Thus, $E_{L}:=E_{Lx}=E_{Ly}$ and $E_{R}:=E_{Rx}=E_{Ry}$. Now, by combining equations (\ref{Meijerx-Taub}), (\ref{Meijery-Taub}) and normalizing (see Appendix A), we obtain the Wigner distribution in the $(x,y,p_{x},p_{y})$-space
\begin{widetext}
\begin{align}\label{N-W-Taub-MWD}
	W_{E_{L}E_{R}}(x,y,p_{x},p_{y})=&\Lambda_{0}\,
	G_{0,4}^{4,0}
	\left(\frac{e^{8x}}{(12\hbar)^{4}}\
	\middle|
	\frac{i}{4\hbar}\left(-p_{x}\pm\sqrt{E_{L}}\right), \frac{i}{4\hbar}\left(p_{x}\pm\sqrt{E_{R}}\right)\right)\nonumber\\
	&\times
	G_{0,4}^{4,0}
	\left(\frac{e^{4y}}{(3\hbar)^{4}}\
	\middle|
	\frac{i}{2\hbar}\left(-p_{y}\pm\sqrt{E_{L}}\right), \frac{i}{2\hbar}\left(p_{y}\pm\sqrt{E_{R}}\right)\right)\,,\nonumber\\\nonumber\\
	\Lambda_{0}:=&
	\frac{\sinh(\pi\left(\sqrt{E_{R}+E_{L}}\right)/4\hbar)}{256\pi^{6}\hbar^{6}\csch(\pi\left(\sqrt{E_{R}+E_{L}}\right)/2\hbar)}\,.
\end{align}
\end{widetext}
Further, as the Wigner distribution behaves as a scalar under linear canonical transformations and, given that the transformation relation between the two distribution functions becomes trivialized \cite{Curtright}, \cite{Kim}, we can revert equation (\ref{N-W-Taub-MWD}) to the original phase space variables $(\Omega,\beta_{+},\pi_{\Omega},\pi_{\beta_{+}})$ by direct substitution. By replacing the transformed coordinates with their original expressions, that is $W_{E_{L}E_{R}}(x,y,p_{x},p_{y})\to W_{E_{L}E_{R}}(\Omega,\beta_{+},\pi_{\Omega},\pi_{+})$, where we used $W_{E_{L}E_{R}}$ for simplicity, thus
\begin{widetext}
\begin{align}\label{TaubSol}
	W_{E_{L}E_{R}}=&
	\Lambda_0\,
	G_{0,4}^{4,0}
	\left(\frac{e^{8(\Omega-2\beta_{+})}}{(12\hbar)^{4}}\
	\middle|
	\frac{i}{4\hbar}\left(\frac{1}{3}(\pi_{\Omega}+2\pi_{+})\pm\sqrt{E_{L}}\right), \frac{i}{4\hbar}\left(-\frac{1}{3}(\pi_{\Omega}+2\pi_{+})\pm\sqrt{E_{R}}\right)\right)\nonumber\\
	&\times
	G_{0,4}^{4,0}
	\left(\frac{e^{4(2\Omega-\beta_{+})}}{(3\hbar)^{4}}\
	\middle|
	\frac{i}{2\hbar}\left(-\frac{1}{3}(2\pi_{\Omega}+\pi_{+})\pm\sqrt{E_{L}}\right), \frac{i}{2\hbar}\left(\frac{1}{3}(2\pi_{\Omega}+\pi_{+})\pm\sqrt{E_{R}}\right)\right)\,,\nonumber\\\nonumber\\
	W=&
	\Lambda\,
	G_{0,4}^{4,0}
	\left(\frac{e^{8(\Omega-2\beta_{+})}}{(12\hbar)^{4}}\
	\middle|
	\pm\frac{i}{4\hbar}\left(\frac{1}{3}(\pi_{\Omega}+2\pi_{+})\pm\sqrt{E}\right)\right)
	G_{0,4}^{4,0}
	\left(\frac{e^{4(2\Omega-\beta_{+})}}{(3\hbar)^{4}}\
	\middle|
	\pm\frac{i}{2\hbar}\left(\frac{1}{3}(2\pi_{\Omega}+\pi_{+})\pm\sqrt{E}\right)\right)\,,\nonumber\\\nonumber\\
	\Lambda:=&
	\frac{\sinh(\pi\sqrt{E}/2\hbar)}{256\pi^{6}\hbar^{6}\csch(\pi\sqrt{E}/\hbar)}\,,
\end{align}
\end{widetext}
where the first and the second identities for $W_{E_{L}E_{R}}$ and $W$ above stand for the non-diagonal and diagonal Wigner distributions, respectively. Also, $\Lambda$ stands for the normalization constant related to the diagonal case, and we have identified $E:=E_{L}=E_{R}$ in such a case. In order to validate the physical consistency of this phase space formulation, the explicit construction of the corresponding standard wave function from this Wigner distribution will be carried out in Section 5.

\subsection{Wigner distribution for Kantowski-Sachs model}
In order to continue we use the Hamiltonian constraint for the Kantowski-Sachs model from equation (\ref{HSH}). Like the previous section, we use a linear canonical transformation to separate the Hamiltonian constraint. We define the new coordinates as $x=2\Omega+\beta$ and $y=\Omega+2\beta$, and again using the generating function of type $F_{2}$, specifically $F_{2}(\Omega, \beta,p_{x},p_{y})=(2\Omega+\beta)p_{x}+(\Omega+2\beta)p_{y}$, we found $\pi_{\Omega}=2p_{x}+p_{y}$ and $\pi_{\beta}=p_{x}+2p_{y}$, moreover, the new momenta are $p_{x}=\frac{1}{3}(2\pi_{\Omega}-\pi_{\beta})$ and $p_{y}=\frac{1}{3}(2\pi_{\beta}-\pi_{\Omega})$. This lead us to a new Hamiltonian constraint
\begin{equation}
	H=-
	\left(3p_{x}^{2}+e^{2x}\right)+
	3p_{y}^{2}=-
	H_{x}+
	H_{y}\approx 0\,.
\end{equation}
The Hamiltonian $H_{x}$ takes a similar form to both Hamiltonians in equation (\ref{Hamiltonian}), thus, we can use a similar development to calculate the Wigner distribution as well as its normalization constant. Consequently we obtain
\begin{widetext}
\begin{equation}\label{Meijerx-KS}
	W_{E_{Lx}E_{Rx}}(x,p_{x})=
	\mathcal{D}\, 
	G_{0,4}^{4,0}
	\left(\frac{e^{4x}}{144\hbar^{4}}\
	\middle|\,
	\frac{i}{2\hbar}\left(-p_{x}\pm\sqrt{\frac{E_{Lx}}{3}}\right),
	\frac{i}{2\hbar}\left(p_{x}\pm\sqrt{\frac{E_{Rx}}{3}}\right)\right)\,.
\end{equation}
\end{widetext}
The Hamiltonian $H_{y}$ takes the form of a free particle system, we consider both star-eigenvalue equations (\ref{Non-diagonal Wigner}) to calculate $W_{E_{Ly}E_{Ry}}(y,p_{y})$. Again, we use $W$ for simplicity, thus
\begin{equation}
	\begin{split}
		H_{y}\star W=&
		3p_{y}^{2}\,W-
		3i\hbar p_{y}\frac{\partial W}{\partial y}-
		\frac{3\hbar^{2}}{4}\frac{\partial^{2}W}{\partial y^{2}}=
		E_{L_{y}}W\,,\\
		W\star H_{y}=&
		3p_{y}^{2}\,W+
		3i\hbar p_{y}\frac{\partial W}{\partial y}-
		\frac{3\hbar^{2}}{4}\frac{\partial^{2}W}{\partial y^{2}}=
		E_{R_{y}}W\,.
	\end{split}
\end{equation}
By subtracting the first line from the second yields
\begin{equation}
	6i\hbar p_{y}\frac{\partial W}{\partial y}=
	(E_{R_{y}}-E_{L_{y}})W\,.
\end{equation}
This differential equation has the solution
\begin{equation}\label{Sol1}
	W=C(p_{y})e^{-\frac{i(E_{R_{y}}-E_{L_{y}})}{6\hbar p_{y}}y},
\end{equation}
where $C(p_{y})$ is a function of $p_{y}$. Moreover, adding both equations and using expression (\ref{Sol1}) above
\begin{equation}
	C(p_{y})
	\left[6p_{y}^{2}+\frac{3\hbar^{2}}{2}\left(\frac{E_{Ry}+E_{Ly}}{6\hbar p_{y}}\right)^{2}-(E_{Ry}-E_{Ly})\right]
	=0\,.
\end{equation}
In the context of distribution theory, an equation of the form $x\cdot f(x)=0$ implies that $f(x)$ must be proportional to the Dirac delta function $\delta(x)$ \cite[p.~1515]{Tannoudji}. Consequently, the distribution $C(p_{y})$ is entirely supported on the roots of the polynomial in the brackets, leading to
\begin{equation}
	C(p_{y})=
	\mathcal{A}\,
	\delta\left(6p{y}^{2}+\frac{3\hbar^{2}}{2}\left(\frac{E_{Ry}-E_{Ly}}{6\hbar p_{y}}\right)^{2}-(E_{Ry}+E_{Ly})\right)\,,
\end{equation}
where $\mathcal{A}$ is a normalization constant. Substituting the explicit form of $C(p_{y})$ back into Equation (\ref{Sol1}) and as the distribution will be performed upon integration we employ the composition property of the Dirac delta function \cite[p.~1515]{Tannoudji}, thus
\beq
	W_{E_{Ly}E_{Ry}}(y,p_{y})
	& = & 
	\frac{\mathcal{A}}{4\sqrt{E_{Ry}E_{Ly}}}
	\sum_{j=1}^{4} |p_{yj}|\,\delta(p_{y} - p_{yj})  \nn\\
	& & \times
	e^{-\frac{i(E_{Ry}-E_{Ly})}{6\hbar p_{y}}y}\,,
	\label{Soly-KS}
\eeq
where the roots are
\begin{equation}
	p_{yj}=
	\pm\frac{\sqrt{E_{R_{y}}}\pm\sqrt{E_{L_{y}}}}{2\sqrt{3}}\,.
\end{equation}
To continue, we combine identities (\ref{Meijerx-KS}) and (\ref{Soly-KS}), and we take $E_{L}:=E_{Lx}=E_{Ly}$ and $E_{R}:=E_{Rx}=E_{Ry}$ to hold the Moyal-Wheeler-DeWitt equation, hence
\begin{widetext}
 \beq
	W_{E_{L}E_{R}}(x,y,p_{x},p_{y})
	& = & 
	\mathcal{B}_{0}
	\sum_{j=1}^{4} |p_{yj}|\,
	\delta(p_{y}-p_{yj})
	e^{-\frac{i(E_{R}-E_{L})}{6\hbar p_{y}}y}\times 
	G_{0,4}^{4,0}
	\left(\frac{e^{4x}}{144\hbar^{4}}\
	\middle|
	\frac{i}{2\hbar}\left(-p_{x}\pm\sqrt{\frac{E_{L}}{3}}\right),
	\frac{i}{2\hbar}\left(p_{x}\pm\sqrt{\frac{E_{R}}{3}}\right)\right)\,,\nonumber\\\nonumber\\
	\mathcal{B}_{0}
	& := & \frac{\mathcal{D}\mathcal{A}}{4\sqrt{E_{L}E_{R}}}=
	\frac{\sinh(\pi(\sqrt{E_{L}}+\sqrt{E_{R}})/2\sqrt{3}\hbar)}{96\pi^{4}\hbar^{4}(E_{L}+E_{R})}\,.
\label{SolKS0}
\eeq
\end{widetext}
To find the normalization constant $\mathcal{A}$ it is straightforward due to the Dirac deltas, where the normalization integral is equal to $\delta(E_{L}-E_{R})$ as discussed in Appendix A, therefore
\begin{equation}
	\mathcal{A}=
	\frac{\sqrt{E_{Ly}E_{Ry}}}{\pi\hbar(E_{Ry}+E_{Ly})}\,.
\end{equation}
In a similar fashion to the Taub model, we can revert this solution to the original phase space variables $(\Omega,\beta,\pi_{\Omega},\pi_{\beta})$ by direct substitution, again, because the Wigner distribution behaves as a scalar under linear canonical, that is $W_{E_{L}E_{R}}(x,y,p_{x},p_{y})\to W_{E_{L}E_{R}}(\Omega,\beta,\pi_{\Omega},\pi_{\beta})$.  From now on, we will use $W_{E_{L}E_{R}}$ for simplicity, thus
\begin{widetext}
\begin{align}
	W_{E_{L}E_{R}}=&
	\mathcal{B}_{0}
	\sum_{j=1}^{4}
	|p_{yj}|\,\delta\left(\frac{1}{3}(2\pi_{\beta}-\pi_{\Omega}) - p_{yj}\right)
	e^{-\frac{i(E_{R}-E_{L})(\Omega+2\beta)}{2\hbar (2\pi_{\beta}-\pi_{\Omega})}}\nonumber\\
	&\times
	G_{0,4}^{4,0}
	\left(\frac{e^{4(2\Omega+\beta)}}{144\hbar^{4}}\
	\middle|
	\frac{i}{2\hbar}\left(-\frac{1}{3}(2\pi_{\Omega}-\pi_{\beta})\pm\sqrt{\frac{E_{L}}{3}}\right),\frac{i}{2\hbar}\left(\frac{1}{3}(2\pi_{\Omega}-\pi_{\beta})\pm\sqrt{\frac{E_{R}}{3}}\right)\right)\,,\nonumber\\\nonumber\\
	W=&
	\mathcal{B}_{1}
	\left\{\delta\left(\frac{1}{3}(2\pi_{\beta}-\pi_{\Omega}) - \sqrt{\frac{E}{3}}\right)+\delta\left(\frac{1}{3}(2\pi_{\beta}-\pi_{\Omega})+\sqrt{\frac{E}{3}}\right)\right\}
	G_{0,4}^{4,0}
	\left(\frac{e^{4(2\Omega+\beta)}}{144\hbar^{4}}\
	\middle|
	\pm\frac{i}{2\hbar}\left(\frac{1}{3}(2\pi_{\Omega}-\pi_{\beta})\pm\sqrt{\frac{E}{3}}\right)\right)\,,\nonumber\\\nonumber\\
	\mathcal{B}_{1}=&
	\frac{\sinh(\pi \sqrt{E}/\sqrt{3}\hbar)}{192\sqrt{3}\pi^{4}\hbar^{4}\sqrt{E}}\,,
\end{align}
\end{widetext}
where the first and second identities for $W_{E_{L}E_{R}}$ and $W$ stand for the non-diagonal and diagonal Wigner distributions, respectively. Note that in the diagonal case for $W$ the relation $E:=E_{L}=E_{R}$ holds and $\mathcal{B}_{1}$ stands for
the normalization constant related to the diagonal case. Following the deparametrization procedure discussed by Simeone \cite{Sime2}, the Hamiltonian constraint surface splits into two disjoint sheets corresponding to positive and negative effective energies. To obtain a physical state that represents a well defined evolutionary branch of the universe, we restrict our analysis strictly to the positive energy sheet. Mathematically, isolating this sector from the full symmetric phase space distribution introduces an overall factor of 2 in the formulation.
\begin{align}\label{SolKS}
		W=&
		\mathcal{B}\,
		\delta\left(\frac{1}{3}(2\pi_{\beta}-\pi_{\Omega})-\sqrt{\frac{E}{3}}\right)  \nn\\
		& \times
		G_{0,4}^{4,0}
		\left(\frac{e^{4(2\Omega+\beta)}}{144\hbar^{4}}\
		\middle|
		\pm\frac{i}{2\hbar}\left(\frac{1}{3}(2\pi_{\Omega}-\pi_{\beta})\pm\sqrt{\frac{E}{3}}\right)\right)\,,\nonumber\\\nonumber\\
		\mathcal{B}=&
		\frac{\sinh(\pi\sqrt{E}/\sqrt{3}\hbar)}{96\sqrt{3}\pi^{4}\hbar^{4}\sqrt{E}}\,.
\end{align}
As for the case of the Taub model, we will validate these results in Section 5 by explicitly constructing the wave function associated with this Wigner distribution.

\section{Wave function in conventional quantum Cosmology}
To establish a physical connection between the quantum description in phase space above and the conventional Quantum Mechanics, we now derive the standard wave functions for both Taub and Kantowski-Sachs cosmological models. By applying the Weyl transform to the diagonal Wigner distributions obtained previously, we can explicitly demonstrate the equivalence of our results with the wave function obtained within the standard Wheeler-DeWitt formulation \cite{Brogaard}. 

\subsection{Wave function for the Taub model}
We will start by considering the diagonal Wigner distribution $W(x,p_{x})$ obtained for the Taub model in equation (\ref{TaubSol}), where we will use the new coordinates and momenta. Inverting equation (\ref{WD-WF}), we can calculate the wave function in terms of the Wigner distribution, that is
\begin{widetext}
\begin{equation}\label{TaubW1}
	\begin{split}
		\psi^{*}(0)\psi(x)=&
		\int_{-\infty}^{\infty}
		W(x/2,p_{x})
		e^{\frac{ip_{x}x}{\hbar}}\,
		dp_{x} = \int_{-\infty}^{\infty}
		G_{0,4}^{4,0}
		\left(\frac{e^{4x}}{(12\hbar)^{4}}\
		\middle|
		\pm\frac{i}{4\hbar}\left(p_{x}\pm\sqrt{E}\right)\right)
		e^{\frac{ip_{x}x}{\hbar}}\, 
		dp_{x}\\
		=&\int_{-\infty}^{\infty}
		\frac{1}{2\pi i}
		\int_{-i\infty}^{i\infty}
		\prod\left\{\Gamma\left[-s\pm\frac{i}{4\hbar}\left(p_{x}\pm\sqrt{E}\right)\right]\right\}
		\left(\frac{e^{4x}}{(12\hbar)^{4}}\right)^{s}
		e^{\frac{ip_{x}x}{\hbar}}\,
		dp_{x}\,,
	\end{split}
\end{equation}
\end{widetext}
where in the last line, once again, we used  the Mellin-Barnes integral definition for the Meijer-G function. The integral over $p_{x}$ is a Fourier transform for the Gamma functions (see Appendix B). Solving that integral, equation (\ref{TaubW1}) becomes
\begin{widetext}
\begin{equation}\label{TaubW2}
	\begin{split}
		\psi^{*}(0)\psi(x)=&
		\frac{4\hbar}{i}
		\int_{-i\infty}^{i\infty}
		\Bigg\{\sum_{n=0}^{\infty}
		\frac{(-1)^{n}}{n!}
		\frac{\Gamma\left[-2s+n+\frac{i\sqrt{E}}{2\hbar}\right]\Gamma\left[-n-\frac{i\sqrt{E}}{2\hbar}\right]}{(\Gamma[-2s+n])^{-1}} e^{-\frac{ix\sqrt{E}}{\hbar}-4x(n-s)}\\
        &		+\sum_{k=0}^{\infty}
		\frac{(-1)^{k}}{k!}
		\frac{\Gamma\left[-2s+k-\frac{i\sqrt{E}}{2\hbar}\right]\Gamma\left[-k+\frac{i\sqrt{E}}{2\hbar}\right]}{(\Gamma[-2s+k])^{-1}}e^{\frac{ix\sqrt{E}}{\hbar}-4x(k-s)}\Bigg\} \left(\frac{e^{4x}}{(12\hbar)^{4}}\right)^{s}\, ds \,.
	\end{split}
\end{equation}
\end{widetext}
The series converges to the hypergeometrical series \cite[p.~1014]{G}, therefore, we can integrate inside the sums. These integrals lead us to obtain Meijer G-functions $G_{0,2}^{2,0}$ \cite[p.~1041]{G}, which can be rewritten in terms of modified Bessel functions of the second kind $K_{\nu}$ \cite[p.~1044]{G} as follows
\begin{widetext}
\begin{equation}
	\begin{split}
		\int_{-i\infty}^{i\infty}
		\Gamma\left[-2s+n+\frac{i\sqrt{E}}{2\hbar}\right]
		\Gamma[-2s+n]
		\left(\frac{e^{8x}}{(12\hbar)^{4}}\right)^{s}\, ds=&
		2\pi i
		\left(\frac{e^{2x}}{12\hbar}\right)^{2n+\frac{i\sqrt{E}}{2\hbar}}
		K_{\frac{i\sqrt{E}}{2\hbar}}\left(\frac{e^{2x}}{6\hbar}\right)\,,\\
		\int_{-i\infty}^{i\infty}\Gamma\left[-2s+n-\frac{i\sqrt{E}}{2\hbar}\right]\Gamma[-2s+n]\left(\frac{e^{8x}}{(12\hbar)^{4}}\right)^{s}\, ds=&2\pi i
		\left(\frac{e^{2x}}{12\hbar}\right)^{2n-\frac{i\sqrt{E}}{2\hbar}}
		K_{\frac{-i\sqrt{E}}{2\hbar}}\left(\frac{e^{2x}}{6\hbar}\right)\,.
	\end{split}
\end{equation}
\end{widetext}
By considering the property $K_{\nu}=K_{-\nu}$, identity (\ref{TaubW2}) becomes
\begin{widetext}
\begin{equation}
	\begin{split}
		\psi^{*}(0)\psi(x)=&
		8\pi\hbar K_{\frac{i\sqrt{E}}{2\hbar}}
		\left(\frac{e^{2x}}{6\hbar}\right)\left\{\left(\frac{1}{12\hbar}\right)^{\frac{i\sqrt{E}}{2\hbar}}
		\sum_{n=0}^{\infty}
		\frac{(-1)^{n}}{n!}
		\Gamma\left[-n-\frac{i\sqrt{E}}{2\hbar}\right]
		\left(\frac{1}{12\hbar}\right)^{2n}\right.\\
		&\left.+
		\left(\frac{1}{12\hbar}\right)^{-\frac{i\sqrt{E}}{2\hbar}}
		\sum_{n=0}^{\infty}
		\frac{(-1)^{n}}{n!}
		\Gamma\left[-n+\frac{i\sqrt{E}}{2\hbar}\right]
		\left(\frac{1}{12\hbar}\right)^{2n}\right\}\,,
	\end{split}
\end{equation}
\end{widetext}
and the involved series converges to modified Bessel function of the first kind $I_{\nu}$ \cite[p.~928]{G} as follows
\begin{widetext}
\begin{equation}
	\begin{split}
		\psi^{*}(0)\psi(x)=&
		8\pi\hbar K_{\frac{i\sqrt{E}}{2\hbar}}\left(\frac{e^{2x}}{6\hbar}\right)
		\left\{I_{\frac{i\sqrt{E}}{2\hbar}}\left(\frac{1}{6\hbar}\right)
		\Gamma\left[1+\frac{i\sqrt{E}}{2\hbar}\right]
		\Gamma\left[-\frac{i\sqrt{E}}{2\hbar}\right]+
		I_{-\frac{i\sqrt{E}}{2\hbar}}\left(\frac{1}{6\hbar}\right)
		\Gamma\left[1-\frac{i\sqrt{E}}{2\hbar}\right]
		\Gamma\left[\frac{i\sqrt{E}}{2\hbar}\right]\right\}\,.
	\end{split}
\end{equation}
\end{widetext}
Using the reflection formula for the Gamma functions $\Gamma[z]\Gamma[1-z]=\pi /\sin(\pi z)$ \cite[p.~603]{Arfken}, and applying the relation between modified Bessel functions of first and second kind $K_{\nu}(z)=\pi(I_{-\nu}-I_{\nu})/2\sin(\pi\nu)$ \cite[p.~937]{G} leads to
\begin{equation}
	\psi^{*}(0)\psi(x)=
	16\pi\hbar K_{\frac{i\sqrt{E}}{2\hbar}}\left(\frac{e^{2x}}{6\hbar}\right)
	K_{\frac{i\sqrt{E}}{2\hbar}}\left(\frac{1}{6\hbar}\right)\,.
\end{equation}
Since the modified Bessel function of the second kind with a purely imaginary order, $K_{i\mu}(z)$, is strictly real-valued for positive real arguments, we can directly identify the components on both sides of the equation. By comparing the $x$-dependent terms and the constant terms, and choosing a trivial global phase, we deduce the normalization constant $|A|^{2}=16\pi\hbar$. Thus, the wave function associated with $W(x,p_{x})$ for the Taub model is
\begin{equation}\label{TxWF}
	\psi(x)=
	\sqrt{16\pi\hbar} K_{\frac{i\sqrt{E}}{2\hbar}}\left(\frac{e^{2x}}{6\hbar}\right)\,.
\end{equation}
To calculate the wave function for $W(y,p_{y})$ we must apply similar arguments as those for $W(x,p_{x})$ above, and therefore, the wave function associated with $W(y,p_{y})$ for the Taub model reads
\begin{equation}\label{TyWF}
	\psi(y)=
	\sqrt{8\pi\hbar} K_{\frac{i\sqrt{E}}{\hbar}}\left(\frac{2e^{y}}{3\hbar}\right)\,.
\end{equation}
Now, bearing in mind identities (\ref{TxWF}) and (\ref{TyWF}) we can write the complete wave function for the Taub model in its canonical coordinates, and taking the constant normalization from the diagonal Wigner distribution from equation (\ref{TaubSol}), we obtain the wave function for the Taub model in the ($x,y$)-space
\begin{equation}
	\begin{split}
		\psi(x,y)=&
		\eta_{0}\,K_{\frac{i\sqrt{E}}{2\hbar}}\left(\frac{e^{2x}}{6\hbar}\right)
		K_{\frac{i\sqrt{E}}{\hbar}}\left(\frac{2e^{y}}{3\hbar}\right)\,\\\\
		\eta_{0}=&
		\frac{\sqrt{2}\sinh(\pi E/2\hbar)}{32\pi^{5}\hbar^{5}\csch(\pi E/\hbar)}.
	\end{split}
\end{equation}
To revert this wave function to its original coordinates we must apply the Dirac transformation function \cite{Blaszak}. For a generating function $F_2(\mathbf{q},\mathbf{p})$ connecting the original coordinates $\mathbf{q}$ to the new momenta $\mathbf{p}$, the integral kernel is determined by the exponentiation of $F_2$, weighted by the square root of the determinant of its mixed second derivatives to ensure unitarity, $|\det(\partial^{2}F_{2}/\partial q_{i}\partial p_{j})|^{1/2}$. Applying this to our specific generating function, the determinant yields exactly $3$. Thus, the explicit transition kernel is given by
\begin{equation}
    \langle\Omega,\beta_{+}|p_{x},p_{y}\rangle=
    \frac{\sqrt{3}}{2\pi\hbar}\exp\left[\frac{i}{\hbar}F_{2}(\Omega,\beta_{+},p_{x}, p_{y}) \right]\,.
\end{equation}
By combining this with the standard projection kernel onto the new coordinate representation \cite[p.~51]{Sakurai} 
\begin{equation}
    \langle x,y|p_{x},p_{y}\rangle=
    \frac{1}{2\pi\hbar}\exp\left[\frac{i}{\hbar}(xp_{x}+yp_{y})\right]\,,
\end{equation}
the complete transition amplitude $\langle x, y | \Omega, \beta_+ \rangle$ is formally constructed via completeness relations. Consequently, the mapping from the original wave function $\psi(\Omega, \beta_+)$ to the new configuration space $\psi(x,y)$ results in the following four-dimensional integral expression
\beq
	\psi(\Omega,\beta_{+})
	& = & 
	\frac{\sqrt{3}}{(2\pi i)^{2}}
	\int_{\mathbb{R}^4}dxdydp_{x}dp_{y}
	\ \left\{\psi(x,y) \right.  \nn\\
	& &  \left.\times e^{-\frac{i}{\hbar}[(\Omega-2\beta_{+}-x)p_{x}+(2\Omega-\beta_{+}-y)p_{y}]}\right\}\,.
\eeq
This integral is straightforward to develop due to Dirac deltas and it shows that to revert canonical coordinates in our case, is just simply substitution with a scale factor, that is $\psi(x,y)\to\sqrt{3}\,\psi(\Omega,\beta_{+})$. Finally, the wave function for Taub model in its original coordinates is
\begin{equation}
	\begin{split}
		\psi(\Omega,\beta_{+})=&
		\eta\,K_{\frac{i\sqrt{E}}{2\hbar}}\left(\frac{e^{2(\Omega-2\beta_{+})}}{6\hbar}\right)
		K_{\frac{i\sqrt{E}}{\hbar}}\left(\frac{2e^{2\Omega-\beta_{+}}}{3\hbar}\right)\,,\\\\
		\eta=&
		\frac{\sqrt{6}\sinh(\pi E/2\hbar)}{32\pi^{5}\hbar^{5}\csch(\pi E/\hbar)}\,.
	\end{split}
\end{equation}
This analytical result constitutes precisely the expected wave function for the quantum Taub minisuperspace model. The functional behavior characterized by the modified Bessel function of the second kind, $K_{\nu}(z)$, is in perfect agreement with the exact solutions obtained via the conventional canonical quantization of the Wheeler-DeWitt equation. This standard behavior was originally established in the foundational work by Moncrief \cite{Mon} regarding the quantum mechanics of the Taub universe, and has been corroborated within distinct time-parameter choices and deparametrization frameworks by Giribet and Simeone \cite{Sime1}. The exact correspondence between our wave function reconstructed from the phase space Wigner distribution and the wave function obtained via conventional quantum cosmology validates the physical consistency and equivalence of the deformation quantization framework for this anisotropic system.

\subsection{Wave function for the Kantowski-Sachs model}
Proceeding as in the previous subsection, we start by considering $W(x,p_{x})$ for the Kantowski-Sachs Model from equation (\ref{SolKS}), and, we use its new coordinates and momenta for simplicity. Thus, following the same development as for the Taub model in the last subsection, its corresponding wave function is
\begin{equation}
	\psi(x)=
	\sqrt{8\pi\hbar} K_{\frac{i\sqrt{E}}{\sqrt{3}\hbar}}\left(\frac{e^{x}}{\sqrt{3}\hbar}\right)\,.
\end{equation}
Further to find the wave function for the case $W(y,p_{y})$ is straighforward as it is a Dirac delta, therefore, its corresponding wave function is
\begin{equation}
	\psi(y)
	=e^{\frac{i\sqrt{E}}{\sqrt{3}\hbar}y}\,.
\end{equation}
Now, we can write the wave function for the Kantowski-Sachs model in its canonical coordinates. Taking the normalization constant from its diagonal Wigner distribution (\ref{SolKS}), we obtain the wave function for the Kantowski-Sachs model in the ($x,y$)-space
\begin{equation}
	\begin{split}
		\psi(x,y)=&
		\xi_{0}\, e^{\frac{i\sqrt{E}}{\sqrt{3}\hbar}y}
		K_{\frac{i\sqrt{E}}{\sqrt{3}\hbar}}\left(\frac{e^{x}}{\sqrt{3}\hbar}\right)\,\\\\
		\xi_{0}=&
		\frac{\sqrt{6}\sinh(\pi E/\sqrt{3}\hbar)}{144\sqrt{(\pi\hbar)^{7}}\sqrt{E}}\,.
	\end{split}
\end{equation}
In order to revert to the original coordinates follows in a complete analogy to the Taub model, where the Dirac transformation function becomes trivialized, that is, to revert canonical coordinates, is just simply substitution with a scale factor, that is $\psi(x,y)\to\sqrt{3}\,\psi(\Omega,\beta)$. Finally, the wave function for Kantowski-Sachs model in its original coordinates is
\begin{equation}
	\begin{split}
		\psi(\Omega,\beta)=&
		\xi\, e^{\frac{i\sqrt{E}}{\sqrt{3}\hbar}(\Omega+2\beta)}
		K_{\frac{i\sqrt{E}}{\sqrt{3}\hbar}}\left(\frac{e^{2\Omega+\beta}}{\sqrt{3}\hbar}\right)\,,\\\\
		\xi=&
		\frac{\sqrt{2}\sinh(\pi E/\sqrt{3}\hbar)}{48\sqrt{(\pi\hbar)^{7}}\sqrt{E}}\,.
	\end{split}
\end{equation}
Analogously, the wave function derived for the Kantowski-Sachs model corresponds exactly to the expected physical state within conventional quantum cosmology. This specific solution coincides precisely with the wave function obtained through the Wheeler-DeWitt equation under canonical deparametrization methods, as explicitly demonstrated by Simeone \cite{Sime2}, where the quantum states are expressed in terms of definite energy solutions of the constraint surface. Furthermore, this characteristic Bessel function profile corresponds to the canonical solutions discussed in the minisuperspace analyses by Cordero, García-Compeán and Turrubiates \cite{Cordero}. Again, we have a exact correspondence between our work and the conventional quantum cosmology, that confirms that deformation quantization formalism successfully captures the complete quantum dynamics of the Kantowski-Sachs model.

\section{Conclusions}
In this paper, we have demonstrated that phase space deformation quantization provides a robust framework for analytically calculating Wigner distributions for anisotropic cosmological models governed by separable Hamiltonians with exponential potentials. By employing the totally symmetric Weyl quantization map, the factor ordering ambiguities commonly encountered in the conventional Wheeler–DeWitt equation are systematically reduced through the adoption of a well-defined ordering prescription. Furthermore, the present approach avoids the formal convergence difficulties typically associated with the Moyal star product, enabling the exact physical states to be obtained in terms of Meijer G-functions. The explicit analytical forms of the Wigner distributions obtained reveal the physical nature of the quantum phase space for each of the anisotropic cosmological models studied here. For the Taub model, the phase space distribution is entirely governed by the product of two Meijer G-functions. This oscillatory structure means that the Wigner distribution inherently takes negative values in certain regions of the phase space. Physically, these negativities are the direct signature of quantum interference between the incoming and outgoing cosmological trajectories as the system bounces against the exponential potential walls generated by its closed spatial curvature. In sharp contrast, the Kantowski-Sachs Wigner distribution exhibits a hybrid nature, characterized by the product of a Meijer G-function and  a Dirac delta distribution, reflecting the mixed dynamical behavior of its Hamiltonian. Specifically, the Dirac delta indicates that the conjugate momentum for the unbounded direction is strictly localized, making this decoupled degree of freedom behave as a free particle. Simultaneously, the Meijer G-function governs the remaining degree of freedom, introducing the characteristic non-classical phase space oscillations and quantum interference associated with the directional collapse. The phase space formulation presented here offers distinct advantages over the conventional quantum cosmology representation. While the standard Wheeler-DeWitt approach provides probability amplitudes strictly confined to the configuration space, the Wigner distribution yields a simultaneous representation of both minisuperspace coordinates and their conjugate momenta. Thus, this formulation could provide a more transparent picture for analyzing the semiclassical limit. Furthermore, the exact analytical distributions obtained could propose a new perspective to explore the behavior of quantum states near the primordial singularity, thanks to these negative regions and oscillations present in the Wigner distributions, leading to a deeper understanding of how quantum interference might alters the classical collapse. Additionally, this formalism presents a natural structure for future works on cosmological decoherence. Since the Wigner distribution represents the phase space symbol of the density matrix, it introduces an ideal mathematical setting to couple the minisuperspace with external interactions. In this context, the quantum to classical transition of the early universe could potentially be studied through the decay of the negative regions inherent to the Meijer G-functions.
\\
\\
A natural step for future research would be to address a mathematical generalization of this framework. In particular, extending the formalism to non-separable Hamiltonians and more general Bianchi classifications could provide a phase space representation for a wider class of physically relevant cosmological systems.

\appendix

\section{Meijer G-function normalization}
In this appendix we normalize the Meijer G-function corresponding to $W{E_{L}E_{R}}(x,p_{x})$ from the Taub model. However, a similar development applies for any other Meijer G-function considered in this paper. To obtain the value of the normalization constant $\mathcal{N}$, we evaluate the integral of the distribution $W_{E_{L}E_{R}}(x,p_{x})$ from equation (\ref{N-W-Taub-MWD}), over the corresponding ($x,p_{x}$)-sector of the whole phase space, thus
\begin{widetext}
\begin{equation}
	\int_{-\infty}^{\infty}
	\int_{-\infty}^{\infty}
	\mathcal{N}
	G_{0,4}^{4,0}
	\left(\frac{e^{8x}}{(12\hbar)^{4}}\
	\middle|\,
	\frac{i}{4\hbar}\left(-p_{x}\pm\sqrt{E_{L}}\right),\frac{i}{4\hbar}\left(p_{x}\pm\sqrt{E_{R}}\right)\right)\, dxdp_{x}\,.
\end{equation}
\end{widetext}
Since the integration variable $p_{x}$ appears within the parameters of the Meijer G-function, we rewrite the function using its Mellin-Barnes integral representation
\beq
	& \frac{\mathcal{N}}{2\pi i}& 
	\int_{-\infty}^{\infty}
	\int_{-\infty}^{\infty}
	\int_{-i\infty}^{i\infty}
	\left\{\prod\frac{\Gamma\left[-s+\frac{i}{4\hbar}\left(- p_{x}\pm\sqrt{E_{L}}\right)\right]}{\Gamma\left[-s+\frac{i}{4\hbar}\left( p_{x}\pm\sqrt{E_{R}}\right)\right]^{-1}}\right\}  \nn\\
	& & \times 
	\left(\frac{e^{8x}}{(12\hbar)^{4}}\right)^{s}\, dsdxdp_{x}\,,
\label{MN1}
\eeq
where the integration contour for $s$ lies along the imaginary axis. It is important to note that product $\Pi$ has not limits, this product runs over all combinations of the $\pm$ signs. The integral over the momenta $p_{x}$ is evaluated using the first Barnes lemma \cite[p.~89]{Andrew} and introducing the change of variable $\alpha=ip_{x}/4\hbar$, therefore
\beq
	& \frac{1}{2\pi i} &
	\int_{-\infty}^{\infty}
	\prod\frac{\Gamma\left[-s+\frac{i}{4\hbar}\left(- p_{x}\pm\sqrt{E_{L}}\right)\right]}
	{\Gamma\left[-s+\frac{i}{4\hbar}\left( p_{x}\pm\sqrt{E_{R}}\right)\right]^{-1}}\, dp_{x}  \nn\\
	& = &
	\frac{\prod\Gamma\left[-2s\pm\frac{i\sqrt{E_{R}}}{4\hbar}\pm\frac{i\sqrt{E_{L}}}{4\hbar}\right]}
	{(-4i\hbar)^{-1}\Gamma[-4s]}\,.
\eeq
Performing the integral over $x$ yields a Dirac delta distribution since $s$ is purely imaginary, thus
\begin{equation}
	\int_{-\infty}^{\infty}
	e^{8xs}\,dx=
	\frac{\pi}{4}\delta(-is)\,.
\end{equation}
Consequently, Equation (\ref{MN1}) can be rewritten as
\begin{equation}\label{MN2}
	\pi\hbar \mathcal{N}
	\int_{-\infty}^{\infty}
	\frac{\prod\Gamma\left[-2iu\pm\frac{i\sqrt{E_{R}}}{4\hbar}\pm\frac{i\sqrt{E_{L}}}{4\hbar}\right]}
	{\Gamma[-4iu]}
	\left(\frac{1}{(12\hbar)^{4}}\right)^{iu}
	\delta(u)\, du \,,
\end{equation}
where we used the change of variable $s=iu$. Evaluating this integral over $u$ requires careful handling of the resulting singularities over the real axis. We displace these singularities of the Gamma functions by considering $u_{s}\rightarrow u_{s}-i\epsilon/2$, and the limit $\epsilon\to 0^{+}$ for $\epsilon>0$. Thus, equation (\ref{MN2}) reduces to
\beq
	& \lim_{\epsilon\to 0^{+}}& 
	\pi\hbar \mathcal{N}
	\int_{-\infty}^{\infty}
	\frac{\prod\Gamma\left[-2iu\pm\frac{i\sqrt{E_{R}}}{4\hbar}\pm\frac{i\sqrt{E_{L}}}{4\hbar}+\epsilon\right]}
	{\Gamma[-4iu+2\epsilon]}
	\nn\\
	& & \times \left(\frac{1}{(12\hbar)^{4}}\right)^{iu}\delta(u)\, du \,.
\eeq
Integration over $u$ using Dirac delta translation property yields
\beq
	& \pi\hbar \mathcal{N} &
	\prod\left\{\Gamma\left[\pm\left(\frac{i\sqrt{E_{R}}}{4\hbar}+\frac{i\sqrt{E_{L}}}{4\hbar}\right)\right]\right\}
	\nn\\
	& & \times
	\lim_{\epsilon\to 0^{+}}
	\frac{\prod\Gamma\left[\pm\left(\frac{i\sqrt{E_{R}}}{4\hbar}-\frac{i\sqrt{E_{L}}}{4\hbar}\right)+\epsilon\right]}
	{\Gamma[2\epsilon]}\,.
\eeq
Again, making use of the reflection formula over the Gamma functions outside the limit and applying the principal part of the Laurent series expansion for the Gamma functions inside the limit we obtain
\beq
	& & \frac{8\pi^{2}\hbar^{2}\mathcal{N}}
	{(\sqrt{E_{R}}+\sqrt{E_{L}})\sinh(\pi(\sqrt{E_{R}}+\sqrt{E_{L}})/4\hbar)}  
	\nn\\
	& & \times
	\lim_{\epsilon\to 0^{+}}
	\frac{\epsilon}{\epsilon^{2}+\frac{(\sqrt{E_{L}}-\sqrt{E_{R}})^{2}}{16\hbar^{2}}}\,.
\eeq
The expression inside the limit is a Cauchy-Lorentz distribution. Applying the limit to this distribution, it thus approaches to a Dirac delta \cite[p.~1515]{Tannoudji}, and therefore we have
\beq
	& & \frac{8\pi^{3}\hbar^{2}\mathcal{N}}
	{(\sqrt{E_{R}}+\sqrt{E_{L}})\sinh(\pi(\sqrt{E_{R}}+\sqrt{E_{L}})/4\hbar)}  \nn\\
	& & \times 
	\delta\left(\frac{\sqrt{E_{L}}+\sqrt{E_{R}}}{4\hbar}\right)\,.
	\label{MN3}
\eeq
Since normalization must be equal to $\delta(E_{Lx}-E_{Rx})$, we rewrite the Dirac delta from equation (\ref{MN3}) in that form leading us to
\beq
	&& \frac{8\pi^{3}\hbar^{2}\mathcal{N}4\hbar(\sqrt{E_{R}}+\sqrt{E_{L}})}
	{(\sqrt{E_{R}}+\sqrt{E_{L}})\sinh(\pi(\sqrt{E_{R}}+\sqrt{E_{L}})/4\hbar)}
	\delta(E_{L}-E_{R})  
	\nn\\
	& & =\delta(E_{L}-E_{R})\,.
\eeq
Finally, the normalization constant is
\begin{equation}
	\mathcal{N}=
	\frac{\sinh(\pi(\sqrt{E_{R}}+\sqrt{E_{L}})/4\hbar)}{32\pi^{3}\hbar^{3}}\,.
\end{equation}
As stated before, a completely analogous development must be applied for $W_{E_{L}E_{R}}(y,p_{y})$ in the Taub model to obtain its own normalization constant, and thus
\begin{equation}
	\mathcal{K}=
	\frac{\sinh(\pi(\sqrt{E_{R}}+\sqrt{E_{L}})/2\hbar)}{8\pi^{3}\hbar^{3}}\,.
\end{equation}
In a similar fashion, the Kantowski-Sachs model presents a Meijer G-function on its own, namely,  $W_{E_{L}E_{R}}(x,p_{x})$ from equation (\ref{SolKS0}). Applying the same development as above for the Taub model, we obtain
\begin{equation}
	\mathcal{D}=\frac{\sinh(\pi(\sqrt{E_{R}}+\sqrt{E_{L}})/2\sqrt{3}\hbar)}{24\pi^{3}\hbar^{3}}\,.
\end{equation}

\section{Fourier Transform for Meijer G-function}
In order to transform our Wigner distribution to its corresponding wave function implies a Fourier transform. In our particular case we will use the diagonal Wigner distribution from the Taub model $W(x,p_{x})$ from equation (\ref{TaubSol}). Also we call $A$ to the integral over $p_{x}$ from equation (\ref{TaubW1})
\begin{equation}
	A=
	\int_{-\infty}^{\infty}
	\prod\left\{\Gamma\left[-s\pm\frac{i}{4\hbar}\left(p_{x}\pm\sqrt{E}\right)\right]\right\}
	e^{\frac{ip_{x}x}{\hbar}}\, dp_{x}\,.
\end{equation}
This function has two infinity sequences of poles on each side of the imaginary axis
\begin{equation}
	\begin{split}
		p_{xm}=&\sqrt{E}-4i\hbar(m-s)\,,\\
		p_{xl}=&-\sqrt{E}-4i\hbar(l-s)\,,\\
		p_{xn}=&-\sqrt{E}+4i\hbar(n-s)\,,\\
		p_{xk}=&\sqrt{E}+4i\hbar(k-s)\,,
	\end{split}
\end{equation}
where we recall that $s$ lies in the imaginary axis. To avoid the use of the principal value of the integral and looking for convergence we displace the poles infinitesimally by adding $i\epsilon/2$, where $\epsilon>0$, that is
\begin{equation}
	\begin{split}
		p_{xm}=&\sqrt{E}-4i\hbar(m-s)+i\epsilon/2\,,\\
		p_{xl}=&-\sqrt{E}-4i\hbar(l-s)+i\epsilon/2\,,\\
		p_{xn}=&-\sqrt{E}+4i\hbar(n-s)+i\epsilon/2\,,\\
		p_{xk}=&\sqrt{E}+4i\hbar(nk-s)+i\epsilon/2\,.
	\end{split}
\end{equation}
Displacing the poles means to change the arguments of the Gamma functions, and therefore, the integral $A$ becomes
\beq
	A & = & \lim_{\epsilon\to 0^{+}}
	\int_{-\infty}^{\infty}
	\frac{\prod\left\{\Gamma\left[-s-\frac{i}{4\hbar}\left(p_{x}\pm\sqrt{E}\right)-\frac{\epsilon}{8\hbar}\right]\right\}}
	{\prod\left\{\left(\Gamma\left[-s+\frac{i}{4\hbar}\left(p_{x}\pm\sqrt{E}\right)+\frac{\epsilon}{8\hbar}\right]\right)^{-1}\right\}}  \nn\\
	& & \times
	e^{\frac{ip_{x}x}{\hbar}}\, dp_{x}\,.
\eeq
To evaluate the integral over the momenta $p_x$, we analytically continue $p_x$ into the complex plane. By restricting our domain to $x>0$, Jordan's lemma allows us to close the integration contour in the upper half-plane. Explicitly, the lemma states that for an integrand containing an exponential factor $e^{iaz}$ with $a>0$, the integral along a large semicircular arc $C_{R}$ vanishes as $R\to\infty$ provided that the remaining algebraic part of the integrand goes to zero uniformly. In our specific case, the positive parameter is $a=x/\hbar>0$, and the product of Gamma functions decays sufficiently at complex infinity, ensuring that the contribution along the infinite semicircular arc vanishes entirely \cite[p.~272]{James}. The integral is then calculated via Cauchy's residue theorem by the residues of the poles of the Gamma functions located in the upper half-plane \cite[p.~234]{James}. Calculating the residues it is noticeable that the term $\epsilon$ vanishes algebraically, without the need to evaluate the limit $\epsilon\to 0^{+}$, thus
\begin{widetext}
\begin{equation}
	\begin{split}
		Res(f(p_{x}),p_{xn})=&
		4\hbar
		\frac{(-1)^{n}}{i(n!)}
		\Gamma\left[-2s+n+\frac{i\sqrt{E}}{2\hbar}\right]
		\Gamma\left[-2s+n\right]
		\Gamma\left[-n-\frac{i\sqrt{E}}{2\hbar}\right]
		e^{-\frac{ix\sqrt{E}}{\hbar}-4x(n-s)}\,,\\
		Res(f(p_{x}),p_{xk})=&
		4\hbar\frac{(-1)^{k}}{i(k!)}
		\Gamma\left[-2s+k-\frac{i\sqrt{E}}{2\hbar}\right]
		\Gamma\left[-2s+k\right]
		\Gamma\left[-k+\frac{i\sqrt{E}}{2\hbar}\right]
		e^{\frac{ix\sqrt{E}}{\hbar}-4x(k-s)}\,,
	\end{split}
\end{equation}
\end{widetext}
where $f(p_{x})$ is indeed the function inside the integral $A$. Finally the Fourier transform we are looking for $A$ is
\begin{widetext}
\begin{equation}
	\begin{split}
		A=&2\pi i
		\left\{
		\sum_{n=0}^{\infty}4\hbar
		\frac{(-1)^{n}}{i(n!)}
		\Gamma\left[-2s+n+\frac{i\sqrt{E}}{2\hbar}\right]
		\Gamma\left[-2s+n\right]
		\Gamma\left[-n-\frac{i\sqrt{E}}{2\hbar}\right]
		e^{-\frac{ix\sqrt{E}}{\hbar}-4x(n-s)}\right.\\
		&\left.+
		\sum_{k=0}^{\infty}4\hbar
		\frac{(-1)^{k}}{i(k!)}
		\Gamma\left[-2s+k-\frac{i\sqrt{E}}{2\hbar}\right]
		\Gamma\left[-2s+k\right]
		\Gamma\left[-k+\frac{i\sqrt{E}}{2\hbar}\right]
		e^{\frac{ix\sqrt{E}}{\hbar}-4x(k-s)}\right\}\,.
	\end{split}
\end{equation}
\end{widetext}
Although this derivation assumes $x>0$ for convergence, the resulting closed-form wave function is completely analytic. Thus, by analytic continuation, this expression represents the globally valid, smooth physical solution for the entire domain $x\in (-\infty, \infty)$ \cite[p.~83]{James}.

\section*{Acknowledgments}
\noindent The authors would like to acknowledge support from SNII SECIHTI (Mexico). JBM acknowledge financial support from Marcos Moshinsky foundation. AM acknowledges financial support from COPOCYT under project 2467 HCDC/2024/SE-02/16 (Convocatoria 2024-03, Fideicomiso 23871). JS was supported by a SECIHTI (Mexico) postgraduate scholarship.

\section*{References}
\bibliographystyle{unsrt}


\end{document}